# INTERPOLATED-DFT-BASED FAST AND ACCURATE AMPLITUDE AND PHASE ESTIMATION FOR THE CONTROL OF POWER

**Józef Borkowski[1), Dariusz Kania[1)**

1) Chair of Electronic and Photonic Metrology, Wroclaw University of Technology, B. Prusa 53/55, 50-317 Wroclaw, Poland
(Jozef.Borkowski@pwr.edu.pl; ✉ Dariusz.Kania@pwr.edu.pl,+48 508 632 287)

**Abstract**

The quality of energy produced in renewable energy systems has to be at the high level specified by respective standards and directives. The estimation accuracy of grid signal parameters is one of the most important factors affecting this quality. This paper presents a method for a very fast and accurate amplitude and phase grid signal estimation using the Fast Fourier Transform procedure and maximum decay sidelobes windows. The most important features of the method are the elimination of the impact associated with the conjugate's component on the results and the straightforward implementation. Moreover, the measurement time is very short - even far less than one period of the grid signal. The influence of harmonics on the results is reduced by using a bandpass prefilter. Even using a 40 dB FIR prefilter for the grid signal with $THD \approx 38\%$, $SNR \approx 53$ dB and a 20-30% slow decay exponential drift the maximum error of the amplitude estimation is approximately 1% and approximately $8.5 \cdot 10^{-2}$ rad of the phase estimation in a real-time DSP system for 512 samples. The errors are smaller by several orders of magnitude for more accurate prefilters.

Keywords: control of power, grid signal, amplitude and phase estimation, renewable energy, interpolated DFT, maximum decay sidelobes windows.

## 1. Introduction

Renewable energy systems allow for electricity production and they have become more and more popular in the U.S., China and EU countries in recent years [1], [2]. One of the most important factors that increase the popularity of such systems is the possibility of any citizen of the country installing one. However, to run the system smoothly (i.e., lack of congestion, high power losses, etc.) and to produce high quality energy, some standards must be fulfilled, e.g., IEC 61727 for photovoltaic systems [3] or IEC 61400 for wind energy systems [4].

One of the most popular renewable energy systems is the photovoltaic system. It consists of three basic components: solar panels, an inverter and a control unit. The inverter is used to convert the DC input signal from the solar panels to the AC output signal, and it is a very important part of the total system. The power control process is based on constantly monitored parameters (frequency, amplitude and phase) of the grid signal. The accuracy of the parameters' estimation has a significant impact on the quality of produced energy and the system reaction time in case of adverse events. The parameters of the grid determine the basis for the algorithm input of the inverter control that is used to govern the switching of the internal transistors [5], [6]. Moreover, the frequency estimate is used, e.g., in the FLL (Frequency Locked Loop) algorithm to synchronize the grid signal and the system output signal [7]; the amplitude estimate is used, e.g., in FD (Fault Detection) algorithms [8] and in under/over voltage protection algorithms [1]; and the phase estimate is used, e.g., in PLL (Phase Locked Loop) algorithms [9] and in the generation of control signals in a controller [10].

This paper presents the development of the grid signal frequency estimation method from [11] for the amplitude and phase estimation method using the FFT procedure and maximum decay sidelobes windows. Many estimation methods of this type allow to obtain very accurate results but in a longer time (more than 1.5 grid signal period in the measurement window) [12], [13]. The method in [11] allows for estimating the frequency in a short time (even far less than one period of the grid signal). The most important features of the method are: the possibility of a short measurement, the high accuracy, the straightforward implementation, minimal computational requirements, and low cost. Moreover, this paper presents an analysis of the harmonic impact on parameter estimation errors and the implementation of methods in a real-time DSP system.

An overview of the spectrum interpolation and the frequency estimation is in [11]. Currently, amplitude and phase estimation methods are also frequently developed. In [14], the author presents a method for the parameters estimation of a multi-sinusoidal signal based on the singular value decomposition method, which are then used to perform the synchronization between the power grid and a distributed generation system. In [15], an estimation method for a stationary and time-varying signals based on the ESPRIT method (Estimation of Signal Parameters via Rotational Invariance Technique) using AWNN (Adaptive Wavelet Neural Network) is proposed; in [16], the phase probability density function is approximated, and then the phase estimation error is minimized using the MSE (Mean Square Error). An interesting analytical solution of the phase estimation, taking into consideration AWGN noise, is presented in [17]; in [18], the method presented is based on the type of a resampling algorithm in the frequency domain, while a recursive technique based on the Gauss–Newton algorithm used for estimation is presented in [19]. An interesting comparison of frequency-domain and time-domain estimation algorithms is presented in [20].

The remainder of this paper is presented as follows. Section 2 presents analytical formulas for amplitude and phase values using the FFT procedure and maximum decay sidelobes windows. In Section 3, the simulation results of the amplitude and phase estimation for the pure signal and the signal in the presence of white Gaussian noise are presented. Section 4 provides the results of the amplitude and phase estimation in response to changing parameters values in the real-time DSP system. Section 5 contains a study on the influence of harmonics on the estimation accuracy and the effect of the prefilter on the achieved results. Section 6 presents the estimation results for the signal that contains harmonics, Gaussian noise and a slow delay exponential component. Finally, the conclusions are presented in Section 7.

## 2. Amplitude and phase estimation method for the case of maximum decay sidelobes windows

This paper presents the parameters estimation method for an $x$(t) signal described in the time domain as follows:

$$x(t) = \sum_{i=1}^{M} A_i \sin\left(2\pi f_i t + \varphi_i\right) \tag{1}$$

Each sinusoidal oscillation is characterized by the amplitude $A_i$, the frequency $f_i$ and the phase $\varphi_i$. Instead of the frequency $f$, it is convenient to determine the normalized frequency $\lambda$ [also referred to as *CiR* (*Cycles in Range*)] with respect to the measurement time $\lambda = f NT$ [bin], where $N$ is the number of $x$(t) signal samples sampled with a frequency of $f_s = 1/T$. The frequency estimation method is presented in [11]; this paper focuses on the presentation of the amplitude and phase estimation method for a sinusoidal signal.

The Discrete-time Fourier Transform (DtFT) of a signal is defined as:

$$X(\lambda) = \sum_{n=0}^{N-1} x_n w_n e^{-j2\pi n\lambda/N} \quad (2)$$

where $x_n = x(nT)$, $n = 0, …, N$-1 and $w_n$ are the time window samples. The spectrum of the primary component can be presented as the sum of the fundamental component characterized by the normalized frequency $\lambda_1$ and the component coupled to a frequency $-\lambda_1$:

$$X(\lambda) = \frac{A_1}{2j} e^{j\varphi_1} W(\lambda - \lambda_1) - \frac{A_1}{2j} e^{-j\varphi_1} W(\lambda + \lambda_1) \quad (3)$$

where

$$W(\lambda) = \sum_{n=0}^{N-1} w_n e^{-j2\pi n\lambda/N} \quad (4)$$

is the DtFT spectrum of the used time window.

In the presented estimation method, maximum decay sidelobes windows (also known as I class Rife-Vincent windows [21] or binomial coefficient windows) are used. They belong to the cosine windows family defined as follows [21]-[23]:

$$w_n = \sum_{h=0}^{P} (-1)^h a_h \cos\frac{2\pi nh}{N}, n = 0,...,N-1 \quad (5)$$

where $P = H$-1 defines number of the used cosines functions, $H$ is the window order, and $a_h$ are window coefficients. Windows used in the method have maximum decay amplitudes of sidelobes from all cosine windows.

The spectrum of the maximum decay sidelobes windows can be approximated for $H > 1$, $\lambda \ll N$, $N \gg 1$ by [23]:

$$W(\lambda) = \frac{D(\lambda)}{P(\lambda)} \quad (6)$$

where:

$$D(\lambda) = \frac{N(2H-2)!}{\pi 2^{2H-2}} \sin(\pi\lambda) e^{-j\pi\lambda} \quad (7)$$

$$P(\lambda) = \lambda \prod_{h=1}^{H-1}(h^2 - \lambda^2) = \frac{\prod_{h=0}^{H}(h^2 - \lambda^2)}{(-\lambda)(H-\lambda)(H+\lambda)} \quad (8)$$

The values of the coefficients $a_h$ for windows with $H \leq 7$, together with the most important parameters, can be found in [11].

The method from [11] allows for determining the normalized frequency $\lambda_1$, and it is based on the relevant three consecutive points of the spectrum $X(\lambda)$ for $\lambda = k$-1, $k$, $k$+1 (values $X_{k-1}$, $X_k$, $X_{k+1}$) around the main lobe, where $k$ is an index of the DFT spectrum $X_k$, i.e. (2) for integer $\lambda$. An important advantage of the method is taking into consideration the effect of a coupled component having a frequency of $-\lambda_1$.

Equations for the amplitude and phase can be determined in an analytical way. Based on (7) and (8), it can be written for $\lambda = k \pm \lambda_1$:

$$D(k \pm \lambda_1) = \frac{N(2H-2)!}{\pi 2^{2H-2}} \sin(\pi(k \pm \lambda_1)) e^{-j\pi(k \pm \lambda_1)} \quad (9)$$

and

$$P(k \pm \lambda_1) = \frac{\prod_{h=0}^{H}\left(h^2 - (k \pm \lambda_1)^2\right)}{\left(-(k \pm \lambda_1)\right)\left(H - (k \pm \lambda_1)\right)\left(H + (k \pm \lambda_1)\right)} \tag{10}$$

Introducing the following notation:

$$F^- = \frac{A_1}{2j} e^{j\varphi_1} D(k - \lambda_1) \tag{11}$$

$$F^+ = -\frac{A_1}{2j} e^{-j\varphi_1} D(k + \lambda_1) \tag{12}$$

taking into account that

$$D(k-1 \pm \lambda_1) = D(k \pm \lambda_1) = D(k+1 \pm \lambda_1) \tag{13}$$

and (6) and (3) for $\lambda = k-1, k, k+1$, three equations are obtained:

$$\begin{cases} \dfrac{F^+}{P(k-1+\lambda_1)} + \dfrac{F^-}{P(k-1-\lambda_1)} - X_{k-1} = 0 \\ \dfrac{F^+}{P(k+\lambda_1)} + \dfrac{F^-}{P(k-\lambda_1)} - X_k = 0 \\ \dfrac{F^+}{P(k+1+\lambda_1)} + \dfrac{F^-}{P(k+1-\lambda_1)} - X_{k+1} = 0 \end{cases} \tag{14}$$

where

$$P(k-1 \pm \lambda_1) = -\frac{(H - k \mp \lambda_1)}{H - 1 + k \pm \lambda_1} P(k \pm \lambda_1) \tag{15}$$

$$P(k+1 \pm \lambda_1) = -\frac{(H + k \pm \lambda_1)}{H - 1 - k \mp \lambda_1} P(k \pm \lambda_1) \tag{16}$$

Based on (11) and (12), the amplitude estimate $\hat{A}_1$ can be obtained, which is burdened with some error (because of the assumptions $\lambda \ll N$, $N \gg 1$ and real signal distortions):

$$\begin{cases} \hat{A}_1 = \dfrac{2jF^-}{e^{j\varphi_1} D(k - \lambda_1)} = \dfrac{e^{-j\varphi_1} 2jF^-}{D(k - \lambda_1)} \\ \hat{A}_1 = -\dfrac{2jF^+}{e^{-j\varphi_1} D(k + \lambda_1)} \end{cases} \Rightarrow \hat{A}_1 = 2\sqrt{\frac{F^- F^+}{D(k + \lambda_1) D(k - \lambda_1)}} \tag{17}$$

where $F^+$ and $F^-$ can be obtained after transformation (14), taking into account two points of the signal spectrum. For example, for the points $X_{k-1}$ and $X_k$:

$$F^- = \frac{P(k-1-\lambda_1)P(k+\lambda_1)}{P(k-1-\lambda_1)P(k+\lambda_1) - P(k-1+\lambda_1)P(k-\lambda_1)} \cdot \left( X_k P(k-\lambda_1) - X_{k-1} \frac{P(k-1+\lambda_1)P(k-\lambda_1)}{P(k+\lambda_1)} \right) \tag{18}$$

$$F^+ = \frac{P(k-1-\lambda_1)P(k+\lambda_1)}{P(k-1-\lambda_1)P(k+\lambda_1) - P(k-1+\lambda_1)P(k-\lambda_1)} \cdot \left( X_{k-1} P(k-1+\lambda_1) - X_k \frac{P(k-1+\lambda_1)P(k-\lambda_1)}{P(k-1+\lambda_1)} \right) \tag{19}$$

Substituting (18) and (19) in (17), the following is obtained for $k=1$ and $H=2$:

$$\hat{A}_1 = 2\frac{e^{j\lambda_1\pi}(1+\lambda_1)}{3N\operatorname{sinc}((\lambda_1-1)\pi)}\cdot\sqrt{-(\lambda_1^2-4)(X_1(\lambda_1+2)-X_0(\lambda_1-1))(X_1(\lambda_1-2)-X_0(\lambda_1+1))} \quad (20)$$

The value of the signal phase can be estimated based on (11) and (12) from:

$$\hat{\varphi}_1 = \arg\left(\frac{2jF^-}{\hat{A}_1 D(k-\lambda_1)}\right) \quad (21)$$

or

$$\hat{\varphi}_1 = \arg\left(-\frac{\hat{A}_1 D(k+\lambda_1)}{2jF^+}\right) \quad (22)$$

For the spectrum points $X_{k-1}$ and $X_k$, (21) for $k=1$ and $H=2$ is:

$$\hat{\varphi}_1 = \arg\left(-\frac{2je^{j\pi(\lambda_1-1)}(\lambda_1-2)}{\hat{A}_1 N\operatorname{sinc}((\lambda_1-1)\pi)(\lambda_1+2+2(\lambda_1-1))}\cdot(\lambda_1 X_1(\lambda_1+1)(\lambda_1+2)-\lambda_1 X_0(\lambda_1^2-1))\right) \quad (23)$$

## 3. Simulation research

Simulation research for the estimation method was performed in the MATLAB environment. The Hanning window ($H = 2$) was used here, which has a good sidelobes suppression and small width of the main lobe. Research included the impact of measurement parameters on the accuracy of the amplitude and phase estimation for the pure sinusoidal signal and for the signal disturbed by Gaussian white noise. The signal frequency was known *a priori*, but, for the practical application, it can be obtained using the method from [11].

At first, the impact of the number of samples $N$ on systematic errors of the amplitude and phase estimation was examined for the pure sinusoidal signal. The signal phase $\varphi_1$ was changed from 0 to $2\pi$ in steps of 0.01 rad. For each *CiR* value, the maximum error of the entire range of $\varphi_1$ was taken. Values of $N$ were changed in the range of $2^5,\ldots, 2^{11}$ (because of the FFT radix-2 algorithm). The value of $k$ was 1 because of the range of tested *CiR* values ($0.1 < CiR < 2$).

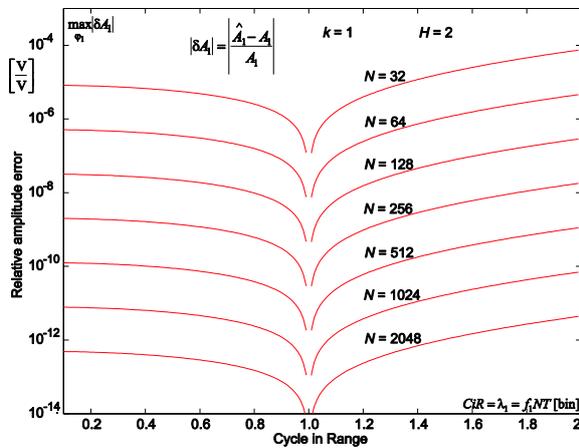
Fig. 1. The systematic amplitude error $\delta A_1$ for the method in Section II: $|\delta A_1|$ is inversely proportional to $N^4$; the smallest error is for *CiR*=1.

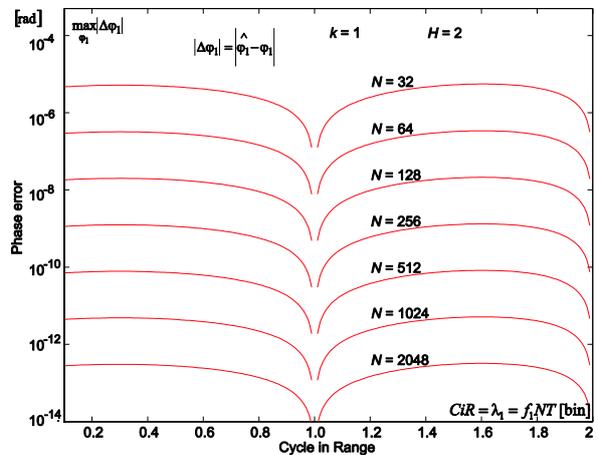
Fig. 2. The systematic phase error $\delta\varphi_1$ for the method in Section II: $|\delta\varphi_1|$ is inversely proportional to $N^4$; the smallest error is for *CiR*=1.

The systematic error of the amplitude and phase estimation method (for the worst phase case) is inversely proportional to $N^4$ (Fig. 1, Fig. 2). Increasing the *N* value at the constant *CiR* value increases the estimation accuracy. The smallest estimation error (global minimum) occurs for *CiR* = 1 − when there is one period of the signal in the measurement window. Figures 1 and 2 differ by the fact that in the first figure there is no local minimum for *CiR* = 2, which is in the second figure due to properties of the method for *H* = 2.

Gaussian white noise that was subsequently added to the signal in the second stage of research was generated using the *randn*() function. The estimation results were compared with corresponding Cramer-Rao bounds (Fig. 3, Fig. 4) [24]. The variance estimator was, in this case *eMSE* (*Empirical Mean Square Error*) assuming that the systematic square error is negligibly small in relation to the variance estimator. The upper value of the *SNR* (*Signal to Noise Ratio*) range corresponds to the actual dynamics of the 16-bit A/D converter. The number of realization was $10^5$, the phase $\varphi_1$ was changed from 0 to $2\pi$ in steps of 0.01 rad and the value of *N* was 512.

For a given *CiR* value, increasing the *SNR* value (decreasing the noise power $\sigma^2$ added to the signal) decreases estimation errors. The square root of *eMSE* errors to the *CRB* bounds was constant regardless of the *SNR* value (Fig. 3, Fig. 4). The ratio value was from approximately 1.76 for *CiR* = 0.7 to approximately 5.25 for *CiR* = 1.5. For *CiR* < 1 errors decrease and for *CiR* > 1 errors increase due to properties of the method for *H* = 2.

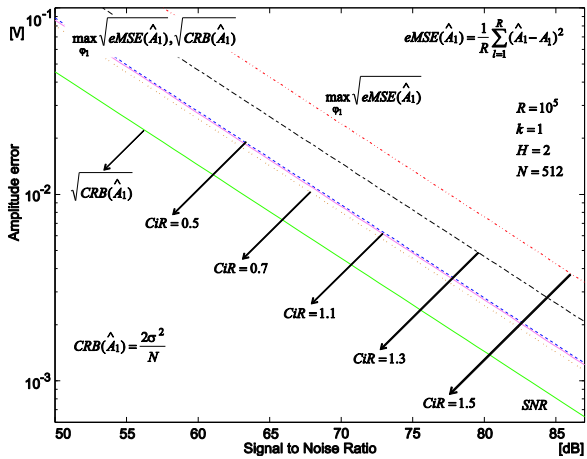
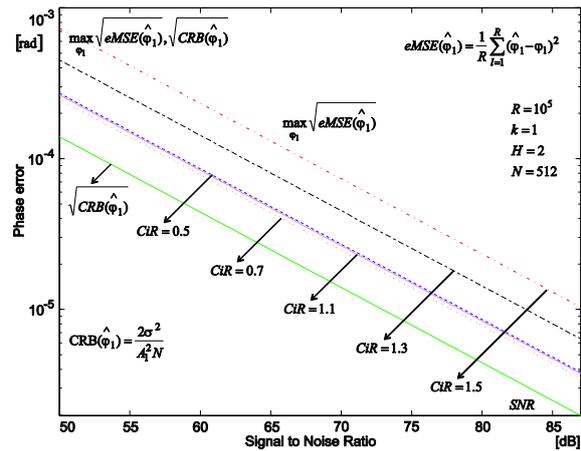

Fig. 3. Statistical properties of the proposed method for estimating the $A_1$ amplitude: the *eMSE* error and the Cramer-Rao bound as a function of *SNR* and for sample *CiR* values.

Fig. 4. Statistical properties of the proposed method for estimating the $\varphi_1$ phase: the *eMSE* error and the Cramer-Rao bound as a function of *SNR* and for sample *CiR* values.

## 4. Real time implementation in a DSP system and reaction time for changing parameters

To confirm the correct functioning of the estimation method and the accuracy of obtained results, the method was implemented in the real-time DSP system of the hardware structure as in [11] (Fig. 5) with the software extension to estimate the amplitude (20) and the phase (23). The test analog signal *x*(*t*) was generated using an arbitrary waveform generator with a 14-bit D/A converter. To convert the *x*(*t*) signal to the digital form, a 16-bit A/D converter at a 24 kHz sampling rate was used. The measurement data processing was performed using a TMS320C6713 floating-point digital signal processor (DSP) with a 225 MHz clock rate. The last *N* samples of the signal were stored in the DSP memory because they were needed to calculate the current estimate. Frequency, amplitude and phase values were updated every four sampling periods (167 μs).

Test signals were generated as recommended by the IEEE Standard for Synchrophasors for Power Systems [25]. They can also be, e.g., test signals used for synchronization between the power grid and the distributed generation system [26]. In the case of the amplitude estimation, the effect of the amplitude's 10% jump at $t = 0$ was examined (Fig. 6a), whereas, in the case of the phase estimation, the effect of the phase 90° jump at $t = 0$ was examined (Fig. 6b).

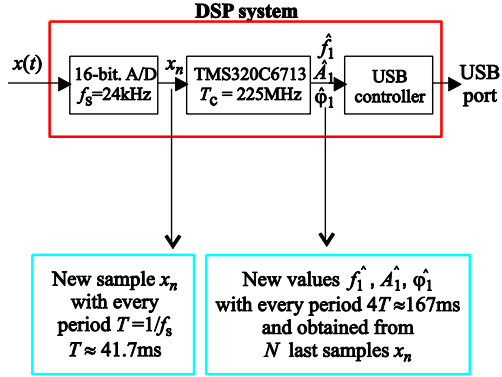

Fig. 5. The real-time measurement system to the signal parameter estimation.

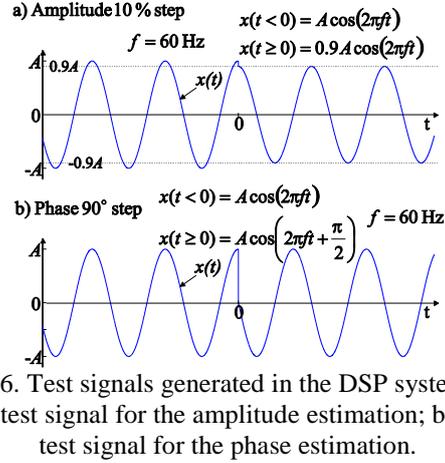

Fig. 6. Test signals generated in the DSP system: a) the test signal for the amplitude estimation; b) the test signal for the phase estimation.

A sudden jump in an estimated value causes the emergence of so-called transient states. The bigger the $N$ value at the constant sampling frequency is (bigger *CiR* value), the longer the transient states are (Fig. 7, Fig. 8). The time required for amplitude estimation with an accuracy of 1% was ca. $2.16 \cdot 10^{-3}$ s (ca. $0.81NT$) for $CiR \approx 0.16$, ca. $4.38 \cdot 10^{-3}$ s (ca. $0.82NT$) for $CiR \approx 0.32$, and ca. $8.38 \cdot 10^{-3}$ s (ca. $0.78NT$) for $CiR \approx 0.64$. Maximum relative estimation errors for the amplitude estimation in the steady state are as follows: for $N = 64$, the error is ca. $3.1 \cdot 10^{-3}$; for $N = 128$, the error is ca. $2.5 \cdot 10^{-3}$; and, for $N = 256$, the error is ca. $1.8 \cdot 10^{-3}$. In the case of the phase estimation, the time required to obtain results with an accuracy of 0.15 rad is ca. $1.39 \cdot 10^{-3}$ s (ca. $0.52NT$) for $CiR \approx 0.16$, ca. $3.27 \cdot 10^{-3}$ s (ca. $0.61NT$) for $CiR \approx 0.32$, and ca. $7.59 \cdot 10^{-3}$ s (ca. $0.71NT$) for $CiR \approx 0.64$ (Fig. 8). Maximum estimation errors in the steady state are as follows: for $N = 64$, the error is ca. $7.4 \cdot 10^{-2}$ rad; for $N = 128$, the error is ca. $6.5 \cdot 10^{-2}$ rad; and, for $N = 256$, the error is ca. $5.5 \cdot 10^{-2}$ rad. The smaller the value of *CiR* is, the bigger the phase spike in the transient state is. The biggest spike is for $N = 64$, but it does not exceed 10% of the actual phase value.

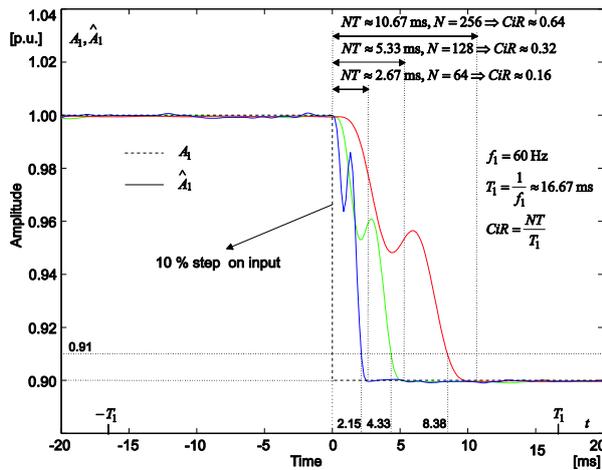

Fig. 7. The effect of the 10% amplitude jump of the tested signal in the amplitude estimation using the proposed method for $N = 64, 128, 256$: the transient time is longer when $N$ is larger, and the 1% accuracy is obtained after ca. $0.8NT$.

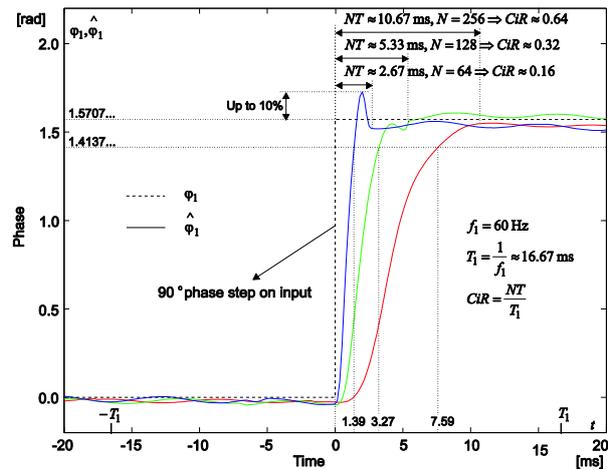

Fig. 8. The effect of the 90° phase jump of the tested signal in the phase estimation using the proposed method for $N = 64, 128, 256$: transient time is longer when $N$ is bigger, and the 0.15 rad accuracy is obtained after ca. $0.5NT$ for $N = 64$, ca. $0.6NT$ for $N = 128$, and ca. $0.7NT$ for $N = 256$.

## 5. The influence of harmonics and the necessity of the pre-filtering

In practice, the grid signal is not a pure sinusoid. It can be disturbed in a deterministic manner by harmonics or in a random manner by white, colored or quantization noise. The presence of harmonics in the grid signal decreases the energy quality. They can have various sources, e.g., work of non-linear loads. The grid signal spectrum contains, except the component (3) corresponding to the grid frequency, $f_1$ additional components of a form similar to (3) but with parameters $A_i$, $f_i$, $\varphi_i$, where $f_i = i \cdot f_1$, $i = 2, 3, \ldots, M$. In practice, the first seven harmonics ($M = 7$) have the greatest impact, and their amplitudes do not exceed 10% of the fundamental component [27].

In simulation research, the effect of the 2$^{nd}$ to the 7$^{th}$ harmonic was determined (Tabs. 1-2) for cases when there was only one harmonic in the tested signal (columns for $i = 2, \ldots, 7$ in Tabs. 1-2) and when there were simultaneously harmonics $i = 2, 3$; $i = 3, 4$, and $i = 2, 3, 4$ (the last three columns in Tabs. 1-2). Amplitudes of harmonics were 10% of the fundamental component, and the estimation error was determined as the maximum amplitude estimation error (Tab. 1) and as the maximum phase error (Tab. 2). The phase $\varphi_1$ of the fundamental component was changed in the full range (0 to $2\pi$ in steps of 0.01 rad), which is necessary because the $\varphi_1$ value has a significant impact on the frequency estimation error [11] as well as the amplitude and the phase estimation error. It is necessary to prefilter the signal to significantly increase the $A_1$ and the $\varphi_1$ estimation accuracy. Tabs. 1-2 show maximum estimation error (for the worst phase $\varphi_1$ case) for the case without prefiltration and using two exemplary band-pass FIR filters designated as A (with a 40 dB band-stop attenuation) and B (with a 60 dB band-stop attenuation). The use of a band-pass filter allows for limiting the influence of harmonics and slow damped disturbances (including the constant component).
The results obtained in Tabs. 1-2 show that estimation errors are strongly dependent on the applied filter and that the use of the A filter allows for achieving the accurate amplitude $A_1$ estimation (error below 1%) and the accurate phase $\varphi_1$ estimation (below 0.001 rad). The use of better filters (e.g., B) allows for improving the estimation accuracy. Here, limitations are maximum error values presented for the pure sinusoid (Figs. 1-2). However, a better filter requires a faster processor in a real-time realization, and usually it increases the system response time.

Table 1. The influence of harmonics 2nd to 7th for the results of the basic component amplitude estimation and its results when using two exemplary band-pass fir pre-filters: A & B.

| CiR | Filter[1] | Amplitude estimation error [%] | | | | | | | | |
|---|---|---|---|---|---|---|---|---|---|---|
| | for[2] $i =$ | 2 | 3 | 4 | 5 | 6 | 7 | 2&3 | 3&4 | 2&3&4 |
| 0.13 | no filter | 15 | 12 | 9.4 | 22 | 38 | 48 | 27 | 19 | 31 |
| | filter A | 0.47 | 0.43 | 0.48 | 0.48 | 0.67 | 0.57 | 0.49 | 0.47 | 0.51 |
| | filter B | 0.041 | 0.045 | 0.041 | 0.041 | 0.059 | 0.066 | 0.047 | 0.046 | 0.048 |
| 0.27 | no filter | 8.9 | 20 | 25 | 13 | 7.6 | 14 | 25 | 45 | 48 |
| | filter A | 0.45 | 0.44 | 0.59 | 0.45 | 0.46 | 0.46 | 0.46 | 0.56 | 0.56 |
| | filter B | 0.039 | 0.049 | 0.046 | 0.039 | 0.041 | 0.046 | 0.051 | 0.058 | 0.059 |
| 0.53 | no filter | 14 | 4.1 | 5.1 | 6.4 | 2.3 | 1.1 | 11 | 8.8 | 5.9 |
| | filter A | 0.47 | 0.42 | 0.45 | 0.42 | 0.42 | 0.42 | 0.46 | 0.45 | 0.5 |
| | filter B | 0.039 | 0.039 | 0.039 | 0.037 | 0.037 | 0.037 | 0.039 | 0.041 | 0.039 |
| 1.07 | no filter | 3.6 | 0.18 | 0.065 | 0.038 | 0.025 | 0.017 | 3.8 | 1.9 | 3.8 |
| | filter A | 0.43 | 0.42 | 0.42 | 0.42 | 0.42 | 0.42 | 0.43 | 0.42 | 0.43 |
| | filter B | 0.037 | 0.037 | 0.037 | 0.037 | 0.037 | 0.037 | 0.038 | 0.037 | 0.038 |

[1] Filter A & B: $F_{stop1}$=10Hz, $F_{pass1}$= 40Hz, $F_{pass2}$= 60Hz, $F_{stop2}$= 90Hz,
 filter A: $A_{stop1}$= 40 dB, $A_{stop2}$= 40 dB, $A_{pass}$= 0.1 dB, order 1686,
 filter B: $A_{stop1}$= 60 dB, $A_{stop2}$= 60 dB, $A_{pass}$= 0.01 dB. order 2738.
[2] All harmonics have amplitude eq. to 10% of the basic component, and the phase eq. 0 rad with respect to the basic component.

Table 2. The influence of harmonics 2nd to 7th for the results
of the basic component phase estimation with exemplary fir pre-filters: A & B.

| *CiR* | Filter[1] for[2] $i =$ | Phase estimation error [rad] | | | | | | | | |
|---|---|---|---|---|---|---|---|---|---|---|
| | | 2 | 3 | 4 | 5 | 6 | 7 | 2&3 | 3&4 | 2&3&4 |
| 0.13 | no filter | $1.5 \times 10^{-1}$ | $1.3 \times 10^{-1}$ | $9.5 \times 10^{-2}$ | $2.3 \times 10^{-1}$ | $3.9 \times 10^{-1}$ | $5.1 \times 10^{-1}$ | $2.8 \times 10^{-1}$ | $1.8 \times 10^{-1}$ | $3.2 \times 10^{-1}$ |
| | filter A | $5.5 \times 10^{-4}$ | $1.6 \times 10^{-4}$ | $6.3 \times 10^{-4}$ | $6.4 \times 10^{-4}$ | $2.5 \times 10^{-3}$ | $1.6 \times 10^{-3}$ | $7.1 \times 10^{-4}$ | $5.9 \times 10^{-4}$ | $8.7 \times 10^{-4}$ |
| | filter B | $2.9 \times 10^{-5}$ | $7.6 \times 10^{-5}$ | $3.3 \times 10^{-5}$ | $3.1 \times 10^{-5}$ | $2.2 \times 10^{-4}$ | $2.9 \times 10^{-4}$ | $1.1 \times 10^{-4}$ | $9.3 \times 10^{-5}$ | $1.2 \times 10^{-4}$ |
| 0.27 | no filter | $8.9 \times 10^{-2}$ | $2.0 \times 10^{-1}$ | $2.6 \times 10^{-1}$ | $1.3 \times 10^{-1}$ | $7.3 \times 10^{-2}$ | $1.4 \times 10^{-1}$ | $2.5 \times 10^{-1}$ | $4.6 \times 10^{-1}$ | $5.0 \times 10^{-1}$ |
| | filter A | $3.2 \times 10^{-4}$ | $2.6 \times 10^{-4}$ | $1.7 \times 10^{-3}$ | $3.6 \times 10^{-4}$ | $4.3 \times 10^{-4}$ | $4.6 \times 10^{-4}$ | $4.9 \times 10^{-4}$ | $1.5 \times 10^{-3}$ | $1.6 \times 10^{-3}$ |
| | filter B | $1.7 \times 10^{-5}$ | $1.2 \times 10^{-4}$ | $9.2 \times 10^{-5}$ | $1.8 \times 10^{-5}$ | $3.8 \times 10^{-5}$ | $8.6 \times 10^{-5}$ | $1.3 \times 10^{-4}$ | $2.1 \times 10^{-4}$ | $2.2 \times 10^{-4}$ |
| 0.53 | no filter | $1.4 \times 10^{-1}$ | $4.0 \times 10^{-2}$ | $5.0 \times 10^{-2}$ | $6.4 \times 10^{-3}$ | $2.2 \times 10^{-3}$ | $1.1 \times 10^{-3}$ | $1.1 \times 10^{-1}$ | $8.7 \times 10^{-2}$ | $6.1 \times 10^{-2}$ |
| | filter A | $5.1 \times 10^{-4}$ | $4.8 \times 10^{-5}$ | $3.5 \times 10^{-4}$ | $1.8 \times 10^{-5}$ | $1.5 \times 10^{-5}$ | $3.6 \times 10^{-6}$ | $4.6 \times 10^{-4}$ | $3.1 \times 10^{-4}$ | $8.1 \times 10^{-4}$ |
| | filter B | $2.6 \times 10^{-5}$ | $2.3 \times 10^{-5}$ | $1.8 \times 10^{-5}$ | $9.1 \times 10^{-7}$ | $1.3 \times 10^{-6}$ | $6.8 \times 10^{-7}$ | $1.2 \times 10^{-5}$ | $4.1 \times 10^{-5}$ | $1.7 \times 10^{-5}$ |
| 1.07 | no filter | $3.6 \times 10^{-2}$ | $1.8 \times 10^{-3}$ | $6.5 \times 10^{-4}$ | $3.8 \times 10^{-4}$ | $2.5 \times 10^{-4}$ | $1.7 \times 10^{-4}$ | $3.8 \times 10^{-2}$ | $2.4 \times 10^{-3}$ | $3.8 \times 10^{-2}$ |
| | filter A | $1.3 \times 10^{-4}$ | $2.3 \times 10^{-6}$ | $4.5 \times 10^{-6}$ | $1.1 \times 10^{-6}$ | $1.6 \times 10^{-6}$ | $5.5 \times 10^{-7}$ | $1.3 \times 10^{-4}$ | $2.6 \times 10^{-6}$ | $1.3 \times 10^{-4}$ |
| | filter B | $6.9 \times 10^{-6}$ | $1.1 \times 10^{-6}$ | $2.4 \times 10^{-7}$ | $5.4 \times 10^{-8}$ | $1.4 \times 10^{-7}$ | $1.1 \times 10^{-7}$ | $7.9 \times 10^{-6}$ | $1.3 \times 10^{-6}$ | $8.1 \times 10^{-6}$ |

[1], [2] See Table 1.

The results in Tabs. 1-2 show also the typical dependence on the *CiR* value (the measurement time). In analogy to the $f_1$ estimation [11], the $A_1$ and the $\varphi_1$ estimation errors usually decrease when the *CiR* is larger in the tested range *CiR* = 0.13, ..., 1.07, such as for the undistorted signal (Figs. 1-2). In the $\varphi_1$ estimation case, this effect can be observed more clearly than in the $A_1$ estimation case. Additionally, when increasing the harmonic number $i$, the $\varphi_1$ estimation error usually decreases (Tab. 2). The influence of interharmonics is similar because there are no significant restrictions for $f_i$. For research in Tabs. 1&2 $f_i$ was $f_i = i \cdot f_1$ because of the greatest significance of this case.

## 6. The real-time implementation with prefiltering

To experimentally verify the effectiveness of the proposed method for the grid signal parameters estimation, the DSP system (Fig. 5) was supplemented in the software layer with the prefiltration using a pass-band FIR filter corresponding to the A filter in Sect. V. The test signal $x(t)$ with frequency $f_1 = 50$ Hz ($\omega = 2\pi f_1$) was distorted in a significant way by four harmonics (*THD* ≈ 38%), white noise with the zero expected value and the standard deviation $\sigma = 0.05$ (that means *SNR* ≈ 53 dB for the 50 Hz component), as well as the exponential component with typical parameters in an industrial load [28]-[31] (Fig. 9). The test signal was generated using the *arbitrary waveform generator* (AWG) with a 14-bit D/A converter. For $N = 512$ (*CiR* ≈ 1.07) and $N = 256$ (*CiR* ≈ 0.53), the signal frequency was estimated by the method from [11], then the amplitude was estimated from (20) and finally the signal phase was estimated from (23). To eliminate harmonics, the DC offset and the slow decay exponential drift of the bandpass FIR filter (the A filter) was applied before the estimation. The lower the *CiR* value is, the shorter the total response time of the estimation method with the filtering is (Fig. 10).

For $N = 512$ (*CiR* ≈ 1.07), the maximum amplitude estimation error was ca. 0.49% without the exponential component and ca. 0.91% after the occurrence of this component, whereas, for $N = 256$ (*CiR* ≈ 0.53), these errors were ca. 0.71% and 0.82%, respectively. For $N = 512$ (*CiR* ≈ 1.07), the maximum phase estimation error was ca. $5.1 \cdot 10^{-2}$ rad without the exponential component and ca. $8.3 \cdot 10^{-2}$ rad after the occurrence of this component, whereas, for $N = 256$ (*CiR* ≈ 0.53), these errors were ca. $6.3 \cdot 10^{-2}$ rad and $7.7 \cdot 10^{-2}$ rad, respectively. The results are more accurate and obtained quicker than the results of the methods from [28]-[31] (including the Least Mean Square Method – LMS), even when the signal frequency is estimated (not known *a priori*, as in [28]-[31]), which is necessary in practice.

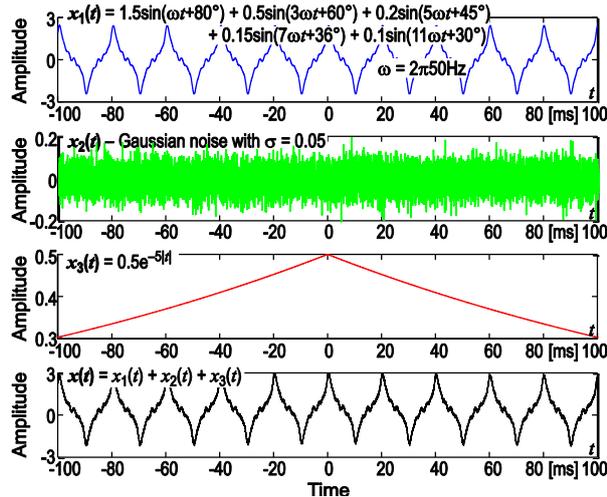

Fig. 9. Grid signal $x(t)$, tested in the final experiment, as a sum of $x_1(t)$ – 50 Hz sinusoid with four harmonics, $x_2(t)$ – Gaussian noise, and $x_3(t)$ – exponential component. All parameter values are the same as in [28]-[31] to allow for comparison of the results.

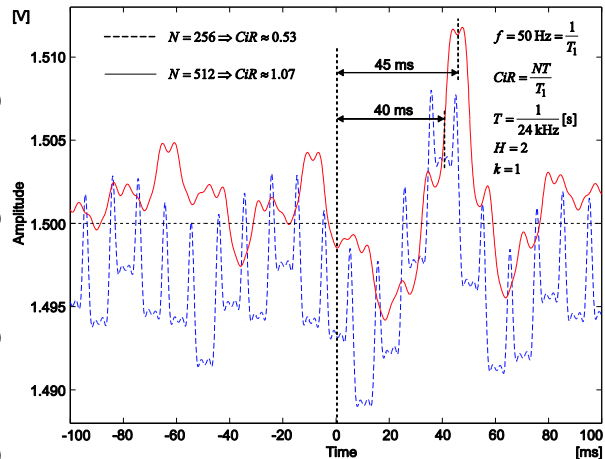

Fig. 10. The proposed estimation method total delay on the example of the amplitude estimation: the marked times, 45 ms (for CiR≈1.07) and 40 ms (for CiR≈0.53), show the delay effect of the exponential part $x_3(t)$ for its maximum value ($t = 0$ and Fig. 9).

The execution time for (20) and (23) is shorter than the execution time for the FFT procedure (Tab. 3) with 6.1% and 6.9% of the processor load, respectively. The prefilter additionally involves the processor in 13.4%, which does not affect significantly the total calculation time. The total load can be reduced by using a smaller number of samples $N$, a prefilter with worse parameters or a processor with a higher clock rate frequency.

Table 3. Execution time on TMS320C6713 in the implemented system.

| Part of the program | Number of cycles $T_c$[1] | | Execution time [µs] | | Processor load [2] [%] | |
|---|---|---|---|---|---|---|
| for $N$ = | 256 | 512 | 256 | 512 | 256 | 512 |
| **Prefiltering** | 5050 | 5050 | 22.4 | 22.4 | 13.4 | 13.4 |
| **FFT algorithm** | 4452 | 11590 | 19.8 | 51.5 | 11.9 | 30.8 |
| $f_1$ **estimation: [11]** | 863 | 863 | 3.8 | 3.8 | 2.3 | 2.3 |
| $A_1$ **estimation: Eq. (20)** | 2290 | 2290 | 10.2 | 10.2 | 6.1 | 6.1 |
| $\varphi_1$ **estimation: Eq. (23)** | 2612 | 2612 | 11.6 | 11.6 | 6.9 | 6.9 |
| Others[3] | 4912 | 5559 | 21.8 | 24.7 | 13.1 | 14.8 |
| Total | 20179 | 27964 | 89.6 | 124.2 | 53.7 | 74.3 |

[1] $T_c = (225\ \text{MHz})^{-1} \approx 4.4\text{ns}$ (TMS320C6713 clock).
[2] Processor load is calculated as the ratio of "Execution time" to the period $4T \approx 167\mu\text{s}$, where $T = 1/f_s \approx 41.7\mu\text{s}$.
[3] Other operations include: reading data from the A/D converter, preparing the data buffer, storing data in memory, etc.

## 7. Conclusions

This paper presents a method to estimate the amplitude and the phase of the grid signal, which is an extension of the frequency estimation method presented in [11]. The FFT procedure and maximum sidelobes windows of $H > 1$ order are used in this method. The possibility of using different windows from the used window family allows for adjusting to the specific problem of signal processing faced.

Computer simulations confirm the correctness of the presented method. The accuracy of the amplitude and the phase estimation for a pure sinusoid depend on the measurement time (*CiR*) and the number of samples *N* collected at this time from the A/D converter (Figs. 1-2) – errors decrease when these values increase (*N* dependency is very strong − inversely proportional to $N^4$). For the difficult case *CiR* = 0.1 (the measurement time is only 10% of the grid signal period, i.e., 1.67 ms for 60 Hz or 2 ms for 50 Hz) and *N* = 2048, the amplitude estimation error is ca. $10^{-12}$ and ca. $10^{-12}$ rad for the phase estimation error as shown in Figs. 1-2. This means that, for a pure sinusoid amplitude and phase, estimation errors, as well as the frequency estimation error [11], are negligible. However, in practice, the grid signal is heavily distorted by subharmonics, harmonics and noise. Moreover, signal parameters may change over time. The algorithm adaptation for changing parameter values in time does not exceed the *NT* value (or equivalent *CiR*) (Figs. 7-8), such as in the case with the change of the frequency and its estimation [11].

The quality of the method for grid signal parameters estimation in the presence of noise is characterized by the ratio of the root square of the variance estimator (here, the *eMSE* error) to *CRB* bounds – the smaller the ratio is, the better the method is, and the impossible limit is 1. For the method presented in this paper, this ratio remains constant regardless of the noise power added to the test signal. For example, for *CiR* = 0.7, this ratio is ca. 1.76, and, for *CiR* = 1.5, it is ca. 5.25. The increase in the *SNR* value (decrease of the noise power added to the signal) causes decreasing estimation errors and *CRB* bounds (Figs. 3-4). A significant source of noise is the quantization noise made by the A/D converter, and, to maintain the high accuracy of the presented method, it is recommended to use the converter with number of bits equal to *b* and not less than 16 [32]. Even for accurate converters ($b \geq 16$), the error caused by the quantization noise is greater than the method error for a pure sinusoid [11], [32].

The biggest impact on the estimation results is caused by the subharmonics and harmonics, as demonstrated by the simulations in Sect. 5 and the experiment described in Sect. 6. Basing the method on the signal spectrum allows for using the prefiltration using universal digital filters. In this paper, band-pass FIR filters were used (A and B filters in Sect. V and the A filter in Sect. 6), but the task of choosing the filter type for different grid signal distortions is, according to the authors, worth further study – the capabilities are large because of the wide variety of different types of digital filters, and the desired objective is to increase the filter accuracy and decrease the delay time brought by the filter.

The comparison with the others methods based on the results obtained by other researchers [28]-[31], including the LMS method (which is known to be very accurate), show that the method based on the interpolated DFT spectrum presented in the paper is a good alternative for LMS and others methods. It also very important that the method uses the same spectrum points to estimate the amplitude, the phase and the frequency [11]. Therefore, the amplitude and phase estimation slightly increases calculation time of the signal parameters (Tab. 3).

The results show that the amplitude and phase estimation method presented in this paper, together with the frequency estimation method of [11], can be successfully applied in control applications of renewable energy systems, especially in the photovoltaic system. The research results of both simulations and the implementation in the real-time DSP system confirm the correctness of the developed methods. The results of the research of Sect. 5 and 6 allow for assessing the influence of individual harmonics in the signal on the parameters estimation accuracy and choosing the prefilter parameters.